\documentstyle[12pt]{article}

\begin{document}

\newpage
\bigskip
\hskip 3.7in\vbox{\baselineskip12pt
\hbox{NSF-ITP-98-086}
\hbox{hep-th/9809022}}

\bigskip\bigskip

\centerline{\large \bf UV/IR Relations in AdS Dynamics}

\bigskip\bigskip

\centerline{\bf Amanda W. Peet\footnote{peet@itp.ucsb.edu} and
Joseph Polchinski\footnote{joep@itp.ucsb.edu}} 
\medskip
\centerline{Institute for Theoretical Physics}
\centerline{University of California}
\centerline{Santa Barbara, CA\ \ 93106-4030}

\bigskip

\begin{abstract}
\baselineskip=16pt
We point out that two distinct distance--energy relations have been
discussed in the AdS/CFT correspondence.  In conformal backgrounds
they differ only in normalization, but in nonconformal backgrounds
they differ in functional form.  We discuss the relation to probe
processes, the holographic principle, and black hole entropies.
\end{abstract}
\newpage
\baselineskip=18pt

An important feature of the recently discovered AdS/CFT duality is a
correspondence between long distances in the AdS space and high
energies in the CFT~\cite{malda,susswitt}.  In fact, at least two  
quite distinct distance--energy relations have been discussed. 
While this point may have been noted implicitly elsewhere, we
believe that it is instructive to discuss it in some detail.
In section~1 we consider various conformally invariant spaces (D3,
M5, and M2).  In section~2 we consider conformally noninvariant
D$p$-brane spaces, where new issues
arise.

\section{Conformal theories}
\subsection{The D3-brane}

For illustration let us consider the near-horizon
geometry of $N$ D3-branes, the 
$AdS_5 \times S_5$ space with string metric
\begin{equation}
ds^2 = \alpha^\prime \Biggl[ {{U^2}\over{g_{\rm YM}{ N^{1/2}}}}
dx_\parallel^2 + \frac{g_{\rm YM}{N^{1/2}}}{U^{2}} 
(dU^2 + U^2 d\Omega_{5}^2) \Biggr] \label{d3met}
\end{equation}
and a constant dilaton $e^{\Phi} = g_{\rm s} = g_{\rm YM}^2$.
We use the conventions of refs.~\cite{malda,IMSY} but omit all
numerical constants.  Susskind and Witten~\cite{susswitt} argue
that imposing an upper cutoff $U$ on the
AdS radius translates into an upper cutoff $E$ on the CFT,
where
\begin{equation}
E = \frac{U}{g_{\rm YM} {N}^{1/2}}\ .
\label{holo}
\end{equation}
One way to obtain this relation is to consider a local change in
the boundary conditions at $U = \infty$.  At radius $U$ the fields
are then perturbed in a region of size\footnote{This follows from
consideration of the wave operator, given in
eq.~(\ref{waveop1}) below.}
\begin{equation}
\delta x_\parallel = \frac{g_{\rm YM} {N}^{1/2}}{U}\ ,
\label{size}
\end{equation}
inverse to the energy~(\ref{holo}).
The relation~(\ref{holo}) leads
to a holographic result for the number of states of the string
theory~\cite{susswitt}: wrapping the
system on a torus of side $L$, the area of surface in Planck units
is
\begin{equation}
\alpha'^{-4}(e^{-2\Phi} \alpha'^4 g_{\rm YM} L^3 U^3 {N}^{1/2})
= L^3 E^3 N^2 \ ,
\end{equation}
which is the indeed the entropy of the cut off gauge
system.

Consider on the other hand a string stretched from the origin
$U=0$ to a probe D3-brane at a radius $U$ as in the original
discussion of Maldacena~\cite{malda}.  The  world-sheet measure
$(G_{tt} G_{UU})^{1/2} =
\alpha'$ just offsets the string tension so that the energy is the
coordinate length
$U$,
\begin{equation}
E = U\ . \label{probe}
\end{equation}
Relations (\ref{holo}) and~(\ref{probe}) are both linear, a
consequence of conformal symmetry.  Although the physics of the CFT
is independent of the scale, these relations are physically
distinct: if one uses the relation~(\ref{probe}) to evaluate the
density of states one obtains the wrong holographic relation,
a fact which led to the current investigation.

There is no contradiction or ambiguity here --- there is no
particular reason that the characteristic energy of probe
processes should be related to the cutoff scale.  
It does, however, seem to contradict a naive renormalization group
interpretation of the AdS/CFT correspondence.  A remark
in ref.~2 helps to resolve this.  The energy
of the stretched string is interpreted in the gauge theory as the
self-energy of a point charge.  This self-energy will be
proportional to $\delta x_\parallel = E^{-1}$ but also to the
effective strength~\cite{maldawilson} $g_{\rm YM} {N}^{1/2}$ of the
Coulomb interaction.  Thus the distinct distance--energy relations
(\ref{holo}) and~(\ref{probe}) are consistent with the single
distance--distance relation~(\ref{size}),
at least when (as here) the effective description is given by low
energy supergravity. 

The energy~(\ref{probe}) is the characteristic gauge theory energy
governing the effective action of a D3-brane probe at a position
$U$ (see however the discussion at the end of this subsection).
Similarly the holographic relation~(\ref{holo}) corresponds to
a probe by one of the massless supergravity fields.  For an $s$-wave
scalar $\psi$ with longitudinal momentum $k$ the wave equation is
\begin{equation}
\left[ -k^2 {{g_{YM}^2 N}\over{U^2}} + 
U^{-3 } \partial_U \left( U^5 \partial_U \right) \right]
\psi=0\ . \label{waveop1}
\end{equation}
By a scaling argument the solution depends only on
$k^2 {g_{YM}^2 N/U^2}$, and so the characteristic radial
dependence of the solution has the holographic
relation~(\ref{holo}) with the energy.  This result is robust ---
it is the scaling such that all terms in the metric are of the
same order --- and holds for all other fields and partial waves
as well.
This holographic relation arises in any context where
the supergravity fields control the physics, including the
temperature/radius relation for the black hole~\cite{susswitt} and
the relation between gauge instanton size and D-instanton
position~\cite{BGKR}.

We should note that the D3-brane probe action is not a
simple Wilsonian action at the scale $U=E$.  Although it appears to
be obtained by integrating out stretched strings of this energy, the
discussion above shows that the size~(\ref{size}) of these 
states in the gauge theory is larger than their Compton wavelength
for large $g_{\rm YM} {N}^{1/2}$;
thus they should not be treated as elementary in loops.  Certain
loop amplitudes, such the celebrated $v^4$ term~\cite{DKPS}, are
protected by supersymmetry and are correctly given, but higher
momentum dependences from the loop graph are incorrect~\cite{DoWa}.

We should also note several other relevant papers, including
refs.~\cite{BKLT,ABKS} which emphasize the differences between
probes, and refs.~[10--16]
which discuss the UV/IR relation from various points of view
including the renormalization group interpretation.

\subsection{The M-branes}

The M5-brane metric
\begin{equation}
ds^2 = l_{11}^2  \Biggl[ {{U^2}\over{ N^{1/3}}}
dx_\parallel^2 + \frac{{N}^{2/3}}{U^{2}} 
(dU^2 + U^2 d\Omega_{4}^2) \Biggr] \label{m5met}
\end{equation}
and the M2-brane metric
\begin{equation}
ds^2 = l_{11}^2  \Biggl[ {{U^2}\over{ N^{2/3}}}
dx_\parallel^2 + \frac{{N}^{1/3}}{U^{2}} 
(dU^2 + U^2 d\Omega_{7}^2) \Biggr] \label{m2met}
\end{equation}
each
lead to the holographic relation 
\begin{equation}
E = \frac{U}{N^{1/2}}\ . \label{mholo}
\end{equation}
This is obtained either from the geodesic equation or the wave
equation for a supergravity probe.  These lead respectively to the
densities of states\footnote{This extension of ref.~\cite{susswitt}
has been noted in refs.~\cite{LiYon,Minic} and in various
unpublished remarks.}
\begin{equation}
S = l_{11}^{-9} (l_{11}^{9} L^5 U^5 N^{1/2}) = L^5 E^5 N^3
\end{equation}
and
\begin{equation}
S = l_{11}^{-9} (l_{11}^{9} L^2 U^2 N^{1/2}) = L^2 E^2 N^{3/2}
\end{equation}
as expected from the nonextremal black hole
entropies~\cite{klebtsey}.

For the M5-brane system, the tension
of an M2-brane stretched to an M5-brane probe at a position
$U$ is simply
$\tau = U^2$, giving the scale
\begin{equation}
E = \tau^{1/2} = U\ .
\end{equation}
This is the characteristic mass gap for an M5-brane probe, and as
in the D3 case it is much larger than the holographic scale.  
For the M2-brane system, we know of no simple picture of the
stretched state that is responsible for the mass gap in the M2-probe
system.  Conformal invariance determines that $E \propto U$ but not
the coefficient.  With the assumption that the scale is
$N$-independent when expressed in terms of the usual D-brane
coordinate $r = U_{\vphantom 1}^{1/2} l_{11}^{3/2}$, one obtains
again the large scale $E = U$.

\section{Nonconformal D$p$-branes}

Although the existence of distinct distance--energy relations does
not lead to an immediate contradiction, it can do so indirectly.
Ref.~\cite{IMSY} analyzes various nonconformal near-horizon
D$p$-brane backgrounds, determining the effective theory governing
the dynamics at various radii.  It is a satisfying result of that
paper that at every radius one can identify a useful effective
theory, but that at no radius is there more than one weakly coupled
effective theory (which would have been a contradiction).  The
effective coupling of the gauge theory depends on energy and so on
the assumed distance--energy relation.  The
stretched-string relation~(\ref{probe}) was used, so it follows that
the analysis is relevant for a D$p$-brane probe but not for a
supergravity probe.  We therefore extend the analysis to the latter
case.

The D$p$-brane near-horizon string metric and dilaton are
\begin{eqnarray}
ds_p^2 &=& \alpha^\prime \Biggl[ {U^{(7-p)/2}\over{g_{\rm YM}{
N^{1/2}}}} dx_\parallel^2 + \frac{g_{\rm YM}{N^{1/2}}}{U^{(7-p)/2}}
(dU^2 + U^2 d\Omega_{8-p}^2) \Biggr]\ , \nonumber\\
e^{\Phi} &=& g_{\rm YM}^2 \Biggl( \frac{U^{7-p} }{g_{\rm
YM}^2 N }\Biggr)^{(p-3)/4}\ .
\label{dpmet}
\end{eqnarray}
Here $g^2_{\rm YM} = g_{\rm s} \alpha'^{(p-3)/2}$.
The wave equation for a massless scalar $\psi$ with angular
momentum $l$, minimally coupled to the Einstein metric $ds'^2_p
= g_{\rm s}^{1/2} e^{-\Phi/2} ds_p^2$, is
\begin{equation}
\Biggl[ - \frac{\partial^2}{\partial U^2} + \frac{(2l + 8 - p)(2l +
6 - p)}{4U^2} + \frac{k^2 g_{\rm YM}^2 N}{U^{7-p}} \Biggr]
U^{(8-p)/2} \psi = 0\ . \label{waveop}
\end{equation}
The energy--distance relations are then
\begin{eqnarray}
\mbox{holographic/supergravity:}&& E = \frac{U^{(5-p)/2}}{g_{\rm YM}
{N^{1/2}}}\ ,
\nonumber\\
\mbox{D$p$-brane:}&& E = U\ .\label{dprelats}
\end{eqnarray}
For the supergravity probe, this is obtained from the wave
equation, or again by requiring that the two terms in
the metric have a common scaling.  For the D$p$-brane probe it is
again determined by the energy of a stretched string.  The
effective gauge coupling is
$g_{\rm eff}^2 = g^2_{\rm YM} N E^{p-3}$ so that
\begin{eqnarray}
\mbox{holographic/supergravity:}&& g_{\rm eff}^2 = [g^2_{\rm YM} N
U^{p-3}]^{(5-p)/2}\ ,
\nonumber\\
\mbox{D$p$-brane:}&& g_{\rm eff}^2 = g^2_{\rm YM} N U^{p-3}\ .
\label{coups}
\end{eqnarray}

In the absence of conformal invariance, the
relations~(\ref{dprelats}) in general no longer have the same
functional form, and so we distinguish several cases.

\subsection{$p \leq 4$}

For $p \leq 4$, energy of the supergravity probe increases with
distance, as in the conformal case.  Moreover, the effective
couplings~(\ref{coups}), though distinct, are related
in a simple way.  The conditions $g_{\rm eff}^2 \ll
1$ are equivalent for the two kinds of probe, and so the
results in ref.~\cite{IMSY} for the effective description of a
D$p$-brane probe apply to the supergravity probe as well.
The effective descriptions given in that paper thus cover the full
range of $U$ for both kinds of probe, with one subtlety of the
$p=1$ case to be discussed below.

It is interesting to extend the analysis of the
holographic principle to these nonconformal cases.  The area of the
surface at radius $U$, in Planck units, is
\begin{eqnarray}
A/G_{\rm N} &=& e^{-2\Phi} L^p \Biggl[{U^{(7-p)/2}\over{g_{\rm YM}
N^{1/2}}}\Biggr]^{p-4} U^{8-p} 
\nonumber\\
&=&
L^p E^{(9-p)/(5-p)} N^2 [g_{\rm YM}^2
N]^{(p-3)/(5-p)}\ . \label{nonexent}
\end{eqnarray}
This is the same as the nonextremal D$p$-brane
entropy~\cite{klebtsey} at the corresponding temperature~$T = E$,
generalizing a result of ref.~\cite{susswitt}.
The interpretation of this entropy has been discussed in
ref.~\cite{klebsuss}.  In particular it has been noted that for
$p=4$ the $E^5 N^3$ behavior agrees with the expectation for a
wrapped M5-brane.  Curiously for $p=1$ the $E^2 N^{3/2}$ behavior
matches that of an M2-brane, even though there is no scale at
which the D1 system is of this form.

Let us add the observation that for $p \leq
3$ the short-distance description is in terms of a dual gauge
theory~\cite{IMSY}, and that the
supergravity result~(\ref{nonexent}) is consistent with this in two
nontrivial respects: the
$N$-dependence agrees with 't Hooft scaling, and the parameters
$\alpha'$ and $g_{\rm s}$ appear only in the combination $g_{\rm
YM}$.  These are not new results, in that they follow from the
scaling of the action as already discussed in ref.~\cite{malda}, 
but it is interesting to contrast this with the understanding
obtained from the correspondence principle~\cite{HoPo}.  The latter
gives a microscopic understanding of the entropy at one
temperature, the boundary between the supergravity and gauge
regimes.  The duality with gauge theory, on the other hand,
restricts the functional form throughout the supergravity regime.
It does not determine the full form --- for that one needs to
understand the exponent of $E$ (the exponent of $g_{\rm YM}$ then
follows by dimensional analysis), which depends on the
interactions.

\subsubsection*{More on $p = 1$}

For $p=1$ there is another issue.\footnote{We would like to thank
N. Itzhaki for bringing this to our attention.  The value $p =1$ is
special because it is the only nonconformal case with odd $p \leq
4$.  For even $p$ the very low energy description is in terms of
$d = 11$ supergravity~\cite{IMSY}.}  In this case the effective
supergravity description breaks down in both the low and high energy
limits.  The high energy description is a gauge theory as discussed
above.  The low energy description is in terms of long free
strings.  The latter description is effective only up to some cutoff
energy, so the effective range in $U$ depends on the relevant
distance--energy relation.   

The following discussion is equivalent to that in ref.~\cite{IMSY},
though we will try to be more explicit about certain points including
the energy--distance relation.  The upper limit of the supergravity
description is
$U = g_{\rm YM} N^{1/2}$ from $g_{\rm eff} = 1$.  For $U < g_{\rm YM}
N^{1/6}$,
$e^\Phi$ is greater than one and the effective string theory is the
$S$-dual of the original.  The local tension of the dual
$\tilde{\rm F}$-string = D-string in terms of the original string
metric is $\tilde \alpha'(U)^{-1} = e^{\Phi(U)} \alpha'^{-1}$.
The condition that the curvature be small compared to this is
\begin{equation}
1 > \tilde \alpha'(U) R = g_{\rm YM}^2 / U^2\ ,
\end{equation}
or $U > g_{\rm YM}$.

At shorter distances the effective
description is in terms of long $\tilde{\rm F}$-strings, the
so-called free orbifold CFT, whose most relevant interaction
is~\cite{DVV}
\begin{equation}
\frac{1}{g_{\rm YM}} \int d^2x\,V_{ij}
\end{equation}
which reconnects the strings $i$ and $j$.  We can verify that the
coefficient here is $N$-independent by considering the large-$g_{\rm
s}$ limit in which $\tilde{\rm F}$-string perturbation theory is a
good description: the coupling in this limit (where $e^{\Phi} =
g_{\rm s}$ and the metric is flat) is
$\tilde g_{\rm s} \tilde \alpha'^{1/2} = g_{\rm s}^{-1/2}
\alpha'^{1/2} = g_{\rm YM}^{-1}$.  Dimensionally the effective
coupling is then $E N^{1/2} / g_{\rm YM}$, in either a thermal
situation ($E = T$) or a scattering process at energy $E$.  The
factor of $N^{1/2}$ is combinatoric: for example, the zeroth order
free energy is of order $N$, while the second order correction in
$V_{ij}$ is of order $N^2$.

It follows that the $\tilde{\rm F}$-string picture is effective for
\begin{equation}
E < E_{\rm DVV} \equiv g_{\rm YM} N^{-1/2}\ .
\end{equation}
(Note that throughout we are measuring energies in terms of the
original coordinates $x_\parallel$.)  Using the holographic relation
$E = U^2 / g_{\rm YM} N^{1/2}$, this becomes $U < g_{\rm YM}$.
Thus, for either supergravity probes or thermal properties the three
descriptions (free CFT, supergravity, gauge theory) cover the full
range of $U$ without overlap.  This was also the conclusion in
ref.~\cite{IMSY}, which in this discussion implicitly used the
holographic distance--energy relation.

On the other hand, the brane probe relation $E = U$ appears to
leave a gap, $U < g_{\rm YM}$ while $E > E_{\rm DVV}$, in which
both the supergravity and free CFT descriptions are
ineffective.  However, we have already emphasized that this
distance--energy relation is much less universal than the holographic
relation.  In the present case, the effective D-string description
of the probe is replaced at these short distances with an effective
$\tilde{\rm F}$-string description, and there appears to be no
relevance to the scale $E = U$ (which would correspond to the loops
of $\tilde{\rm D}$-strings).

\subsection{$p = 5$}

This case has recently been discussed in ref.~\cite{ABKS}, and
our analysis will overlap with that paper.  Ref.~\cite{BarRab}
also notes that holography appears to break down for $p \geq 5$.

For $p=5$ the holographic/supergravity relation between energy and
radius degenerates: a supergravity field of given energy does not
probe a characteristic radius.  Rather, an on-shell field
propagates along the throat indefinitely.  In the small-curvature
region $U > (g_{\rm YM} N^{1/2})^{-1/2}$
the supergravity
description is valid (though with an $S$-duality crossover 
in terms of the underlying string theory~\cite{IMSY}).  There are
two solutions in this region, $\chi \propto U^{\beta_\pm}$ with
\begin{equation}
(\beta + 3/2)(\beta+ 1/2) = (l + 3/2)(l+ 1/2) + k^2 g_{\rm YM}^2
N\ . \label{roots}
\end{equation}
The reflection coefficient depends on physics in the small-$U$
high-curvature region.  For small $k^2 g_{\rm YM}^2 N$ this will
be determined by a weakly coupled gauge theory on the D5-branes.
Note that also in this regime the energy-dependence from the
supergravity solution~(\ref{roots}) is weak.  For large
$k^2 g_{\rm YM}^2 N$, there is no effective theory for the
high-curvature region.  Here, however, the energy-dependence in
the supergravity region is strong and one might expect that this
is the dominant effect.  Thus there are effective
descriptions for all energies.

Let us make a further remark about the holographic idea in this
context.\footnote{Similar observations have been made by Simon
Ross.}  Consider an upper cutoff $U_0$, and imagine a perturbation
of the boundary condition which is local in $x_\parallel$.  At $U <
U_0$ it follows from the wave operator~(\ref{waveop}) that the
fields are perturbed in a region of size
\begin{equation}
\Delta x_\parallel \sim g_{\rm YM} N^{1/2} \ln \frac{U_0}{U}\ .
\label{spread}
\end{equation}
In order to relate the boundary field perturbation to a
renormalized local operator in the boundary quantum
theory~\cite{witten,GKP} we must take $U_0$ to infinity while
holding the fields fixed at a `renormalization scale' $U$.  Unlike
the AdS case (and the $p \leq 4$ branes) this is not possible:
according to~(\ref{spread}) the fields at $U$ spread indefinitely.
Thus renormalized local operators do not exist.\footnote{A similar
pathology occurs in another context in which the asymptotic geometry
is flat
\cite{GHP}.}  One can also see this in momentum space: the generic
solution grows as
\begin{equation}
U^{\beta_+} \sim U^{(g^2_{\rm YM} N k^2 )^{1/2}}
\end{equation}
at large $U$.  The boundary data must scale in this way in order to
produce a renormalized result; however, this has no Fourier
transform.  For $p \leq 4$ there is no such momentum-dependent
renormalization.  This may shed light on the point made in
ref.~\cite{ABKS}, that the boundary theory cannot be a normal field
theory.  Fourier modes of quantum fields exist, but not the local
fields themselves.

\subsection{$p = 6$}

In this case the radius probed by a supergravity field varies
inversely with the energy, in contrast to the usual expectation. 
Correspondingly, the effective gauge couplings~(\ref{coups}) for
the two probes have opposite weak coupling regimes.  When
$g^2_{\rm YM} N E^3$ is large there is no effective theory: the
supergravity background is highly curved and also the gauge theory
is strongly coupled.  When $g^2_{\rm YM} N E^3$ is small there is
the opposite problem: both the gauge and supergravity descriptions
are weakly coupled.

This is the classic contradiction, whose
avoidance in most circumstance is one of the striking evidences
for duality~\cite{witusc}: the existence of distinct weakly coupled
descriptions would surely lead to inconsistent results for some
observables. 
In this case, however, we believe that the resolution is rather
prosaic.  Namely, the low energy Hilbert space has two sectors,
a gauge theory which describes D6-brane probes close to the origin
and a supergravity theory which describes supergravity probes far
away.  These sectors are isolated from one another, being
respectively at
$U^3$ less than and greater than $g^{-2}_{\rm YM} N^{-1}$.  
The unusual behavior of this example, not seen in any of the
others, is likely related to the nonexistence of an underlying
field theory~\cite{IMSY}.

\section*{Acknowledgments} 

We have learned that Sunny Itzhaki has also
considered many of the issues in this paper.  We would like to thank
him, and also Steven Gubser, Aki Hashimoto and Simon Ross, for
discussions.  A. W. P. would like to thank the Aspen Center for
Physics for hospitality during the later stages of this work.  This
work was supported in part by NSF grants PHY94-07194 and PHY97-22022.

\end{document}